\documentclass[doublecol,linenumbers]{epl2} 
\usepackage{microtype}
\usepackage{amsmath}
\usepackage{amssymb}
\usepackage{graphicx}
\usepackage{wasysym}
\usepackage{color}

\title{Fluctuation relations for anisotropic systems}

\author{R. Villavicencio-Sanchez\inst{1} \and R. J. Harris\inst{1} \and H. Touchette\inst{2,3}}

\institute{                    
  \inst{1} School of Mathematical Sciences, Queen Mary University of London - London E1 4NS, UK\\
  \inst{2} National Institute for Theoretical Physics (NITheP) - Stellenbosch 7600, South Africa\\
  \inst{3} Institute of Theoretical Physics, University of Stellenbosch - Stellenbosch 7600, South Africa
}
\pacs{05.40.-a}{Fluctuation phenomena, random processes, noise, and Brownian motion}
\pacs{05.70.Ln}{Non-equilibrium and irreversible thermodynamics}
\pacs{02.50.-r}{Probability theory, stochastic processes, and statistics}

\abstract{Currents of particles or energy in driven non-equilibrium steady states are known to satisfy certain symmetries, referred to as fluctuation relations, determining the ratio of the probabilities of positive fluctuations to negative ones. A generalization of these fluctuation relations has been proposed recently for extended non-equilibrium systems of dimension greater than one, assuming, crucially, that they are isotropic [P.~I.\ Hurtado, C.\ P\'erez-Espigares, J.~J.\ del Pozo, and P.~L.\ Garrido, Proc.\ Nat.\ Acad.\ Sci.\ (USA) \textbf{108}, 7704 (2011)]. Here we relax this assumption and derive a fluctuation relation for $d$-dimensional systems having anisotropic bulk driving rates. We test the validity of this anisotropic fluctuation relation by calculating the particle current fluctuations in the 2-$d$ anisotropic zero-range process, using both exact and fluctuating hydrodynamic approaches.}

\begin{document}

\newcommand{\mb}[1]{\boldsymbol #1}
\newcommand{\sigmas}{ \Sigma}
\newcommand{\ra}{\rightarrow}

\maketitle

\section{Introduction}

Fluctuations play an important role at small and mesoscopic scales, for example in nano-devices, chemical reactions, and molecular motors~\cite{klages2013,bustamante2005,julicher1997,ritort2008}. Depending on the properties of the medium and applied forces considered (e.g., the shape of a trapping potential or the spatial distribution of a reactant), such fluctuations may be isotropic or anisotropic and often show certain symmetry properties, such as the \emph{Gallavotti-Cohen fluctuation relation} (GCFR), which has been the subject of considerable theoretical and experimental study~\cite{maes2003b,kurchan2007,harris2007,seifert2012,ritort2003a}. The GCFR applies to scalar observables of driven non-equilibrium systems and implies the following relation between positive and negative fluctuations integrated over a time $t$:
\begin{equation}
\lim_{t\rightarrow\infty} -\frac{1}{t}\log\frac{P(-A,t)}{P(A,t)}=c A.
\label{eqgcfr1}
\end{equation}
Here $P(A,t)$ denotes the probability density function (pdf) of the time-averaged observable $A$ and $c$ is a time-independent constant. This relation has been derived for many non-equilibrium observables, including the entropy production of chaotic systems, integrated currents in interacting particle models, and work- or heat-like quantities defined in the context of driven Langevin equations~\cite{evans1993,gallavotti1995,kurchan1998,lebowitz1999}. The GCFR has also been verified experimentally, e.g., in turbulent fluids~\cite{ciliberto2004} and for manipulated Brownian particles~\cite{wang2002,ciliberto2010}.

Recently, Hurtado \textit{et al.}~\cite{hurtado2011} have proposed a generalization of the GCFR, called the \emph{isometric fluctuation relation} (IFR), in an effort to uncover new fluctuation symmetries for higher-dimensional systems. Instead of considering positive and negative fluctuations of scalar observables, their IFR focuses on the global current vector $\mb{J}$ of $d$-dimensional non-equilibrium systems and implies that any two currents $\mb{J}'$ and $\mb{J}$ of equal magnitude, $|\mb{J}'|=|\mb{J}|$, obey the following relation:
\begin{equation}
\lim_{tL^d\ra\infty}-\frac{1}{tL^d}\log \frac{P(\mb{J}',t)}{P(\mb{J},t)}=\mb{E}\cdot (\mb{J}-\mb{J}'),
\label{eqifr1}
\end{equation}
where $L^d$ is the volume of the system and $\mb{E}$ is a $d$-dimensional current-independent field conjugate to $\mb{J}$. This relation can be derived from the hydrodynamic fluctuation theory and has been shown to hold so far for a number of important non-equilibrium models, including the boundary-driven 2-$d$ Kipnis-Marchioro-Presutti (KMP) process and a hard-disk fluid model~\cite{hurtado2011}.

Crucially, both the derivation and the application of the IFR rely on the systems of interest being isotropic. Our goal here is to remove this assumption so as to derive a fluctuation relation similar to (\ref{eqifr1}) but which applies to more general systems having anisotropic diffusive dynamics. As a test of this anisotropic fluctuation relation (AFR), we consider the 2-$d$ zero-range process on a square lattice with different hopping rates in $x$- and $y$-directions. We obtain the current fluctuations in this model from the hydrodynamic fluctuation theory as well as exactly from the microscopic definition of the process for system sizes up to $10^5\times 10^5$ sites, which is much larger than currently accessible in numerical simulations. This allows us to determine in a precise way the regime of current fluctuations for which the AFR effectively describes the fluctuation symmetries of extended systems.

\section{Hydrodynamic formalism and IFR}

We study diffusive lattice gases evolving on a $d$-dimensional (hypercubic) lattice of side $L$. In the macroscopic scaling limit, these systems are described, following the hydrodynamic fluctuation theory~\cite{bertini2002,bertini2005a,bertini2006,bodineau2004}, by a local particle density $\rho(\mb{r},t)$, with $\mb{r} \in \Omega=[0,1]^d$, and a local current
\begin{equation}
	\mb{j}(\mb{r},t)=-D(\rho)\nabla\rho(\mb{r},t)+\mb{\xi}(\mb{r},t) \label{MacroJ}.
\end{equation}
Density boundary conditions account physically for the interaction with reservoirs while mass conservation imposes the continuity equation
\begin{equation}
\partial_t \rho(\mb{r},t)=-\nabla\cdot \mb{j}(\mb{r},t).
\label{eqcont1}
\end{equation}
The local current $\mb{j}(\mb{r},t)$ is composed of two parts: a deterministic drift with a density-dependent \emph{diffusivity} $D(\rho)$, representing the hydrodynamic (noiseless) evolution of the model, and a random noise $\mb{\xi}(\mb{r},t)$, accounting for the fluctuations of the model around its hydrodynamic behaviour. This noise is assumed to be a space-time white noise with covariance $L^{-d}\sigma(\rho)\delta(\mb{r'}-\mb{r})\delta(t'-t)$, where $\sigma(\rho)$ is the density-dependent \emph{mobility}. To allow for any anisotropy in the system, $\sigma(\rho)$ and $D(\rho)$ are here taken to be $d\times d$ matrices rather than scalar functions.

The non-equilibrium state of the model is characterized by the \emph{global current} averaged over time $t$,
\begin{equation}
\mb{J}=\frac{1}{t}\int_{0}^{t}d\tau\int_{\Omega}d\mb{r}'\mb{j}(\mb{r'},\tau). \label{global-current}
\end{equation}
For some choices of boundary conditions and matrices $D$ and $\sigma$, $\mb{J}$ converges in the long-time limit, $tL^d\rightarrow\infty$, to a typical value, corresponding to the hydrodynamic current. Here we are interested in fluctuations of $\mb{J}$ about this limit and in any symmetries satisfied by its pdf $P(\mb{J},t)$. In most cases, this pdf has an exponential form in $t$ and $L^d$,
\begin{equation}
P(\mb{J},t)=\exp\left(- tL^d \hat{e}(\mb{J})+o\left(tL^d\right)\right),
\label{LDP}
\end{equation}
which is referred to as a large deviation principle~\cite{touchette2009}. The rate function $\hat{e}(\mb{J})$ characterizes the speed at which $P(\mb{J},t)$ converges to its typical value, and so quantifies the asymptotic probability of rare current fluctuations.

From a microscopic point of view, current and density fluctuations are linked. In the hydrodynamic limit, it can be shown that a given value of the global current $\mb{J}$ is overwhelming likely to be realized by particular spatio-temporal profiles of the local density and current, referred to as \textit{optimal} profiles. These optimal profiles and the corresponding rate function $\hat{e}(\mb{J})$ are obtained within a path integral formalism by an asymptotically exact ``saddle-point'' calculation over all realizations of the noise which yield local currents~(\ref{MacroJ}) consistent with the desired global value $\mb{J}$.

Following this picture, the IFR of (\ref{eqifr1}) can be derived, as in~\cite{hurtado2011}, under the following assumptions: (\emph{i}) isotropic diffusivity and mobility (i.e., $D(\rho)$ and $\sigma(\rho)$ proportional to the identity matrix), (\emph{ii}) time-reversible dynamics with local detailed balance, (\emph{iii}) time-independent optimal profiles for both current and density, and (\emph{iv}) space independent optimal current profiles (i.e., homogeneous local current). This last assumption can be omitted, in fact, but the resulting generalised IFR is for \emph{local} rotations of divergenceless current profiles and does not have the same simple structure as~(\ref{eqifr1}).

Underlying the IFR is the remarkable property that the optimal density profile is the same for all currents $\mb{J}$ on a \emph{circle} of given radius around the origin. Our contribution is to look for a similar relation to~(\ref{eqifr1}) but, significantly, without assuming the isotropic condition (\emph{i}).

\section{Anisotropic fluctuation relation}

We now aim to determine which currents can be related via a fluctuation relation of the same type as (\ref{eqifr1}). We assume that the diffusive system has open boundary conditions in the $x$-direction, without loss of generality, and periodic boundary conditions in the other $(d-1)$ directions. From the macroscopic fluctuation theory outlined above, the rate function of $\mb{J}$ is obtained from the following optimization problem~\cite{bertini2002,bertini2005a,bertini2006,bodineau2004}:
\begin{equation}
	\hat e(\mb{J})=\min\limits_{\rho,\mb{j}}\frac{1}{t}\int_0^t d\tau\int_{\Omega} d\mb{r}\mathcal{L}\left( \tau,\mb{r},\rho,\nabla\rho \right),
\label{eqmin1}
\end{equation}
involving the Lagrangian
\begin{equation}
	\mathcal{L}\left( \tau,\mb{r},\rho,\nabla\rho \right)=\frac{\left( \mb{j}(\mb{r},\tau)+D\nabla\rho \right)^{T} \sigmas \left( \mb{j}(\mb{r},\tau)+D\nabla\rho \right) }{4}.
\label{RF-functional}
\end{equation}
Here $\sigmas$ is the diagonal matrix with elements $\sigmas_{kk}=\left(\sigma_k\right)^{-1}$, where $\sigma_{k}=\Lambda_k^{-1}f(\rho)$ is the mobility in the $k$th direction and $\Lambda$ is a diagonal matrix independent of the density. Similarly, the diffusivity $D(\rho)$ is a diagonal matrix with elements $D_k=\Delta_kg(\rho)$. This factorised form of diffusivity and mobility matrices encompasses a large class of physical systems where an interaction process takes place at different rates in different directions. The local density and current solving the minimization (\ref{eqmin1}) with the continuity equation (\ref{eqcont1}) and boundary conditions are the optimal profiles mentioned above.

The constrained optimization (\ref{eqmin1}) is very difficult to solve in general. However, following~\cite{hurtado2011}, it can be simplified under hypothesis (\emph{iii}) and (\emph{iv}) above to obtain
\begin{equation}
	\hat e(\mb{J})=\min\limits_{\rho}\frac{1}{4}\int d\mb{r}\left( \mb{J}+D\nabla \rho \right)^{T} \sigmas \left( \mb{J}+D\nabla \rho \right).\label{SimplRF}
\end{equation}
Furthermore, assumption (\emph{ii}) for diffusive gases means
\begin{equation}
\frac{\delta}{\delta\rho}\int_\Omega d\mb{r}\sigmas D\nabla\rho=0,\label{timerev}
\end{equation}
which for the minimization of~(\ref{SimplRF}) implies
\begin{equation}
\frac{\delta}{\delta\rho}\int_\Omega d\mb{rJ}^T\sigmas\mb{J}=-\frac{\delta}{\delta\rho}\int_\Omega d\mb{r}\left(D\nabla\rho\right)^T\sigmas\left(D\nabla\rho\right)
\label{MinDens1}
\end{equation}
and we observe that, significantly, the solutions depend only on $\mb{J}^T\sigmas\mb{J}$. Defining the constant field
\begin{equation}
\mb{E}=\frac{1}{2}\int_\Omega d\mb{r}\sigmas D\nabla\rho,
\label{field}
\end{equation}
and taking the difference of rate functions corresponding to the left hand side of~(\ref{eqifr1}), we then obtain the fluctuation relation
\begin{equation}
\hat e(\mb{J})-\hat e(\mb{J'})=\mb{E} \cdot \left(\mb{J}'-\mb{J}\right)
\label{FinalFR}
\end{equation}
for global currents satisfying
\begin{equation}
\mb{J}^{T}\Lambda\mb{J}=\mb{J}'^{T}\Lambda\mb{J}'.
\label{eqell1}
\end{equation}
The two equations above define our anisotropic generalization of the IFR, called AFR, showing that the relation of Eq.~(\ref{eqifr1}) is now valid for currents on \emph{ellipses} determined by (\ref{eqell1}), rather than the circles obtained in \cite{hurtado2011} for $\sigmas$ and $D$ proportional to the identity matrix. Similarly, generalizing the underlying structure, we also see from Eq.~(\ref{MinDens1}) that currents on a given ellipse arise from the same optimal density profile. In other words, optimal density profiles are invariant on ellipses, with principal axes determined by the anisotropy, as encoded in the matrix $\Lambda$.

\section{Zero-range process}

\begin{figure}[t]
	\begin{center}
		\includegraphics[width=2.8in]{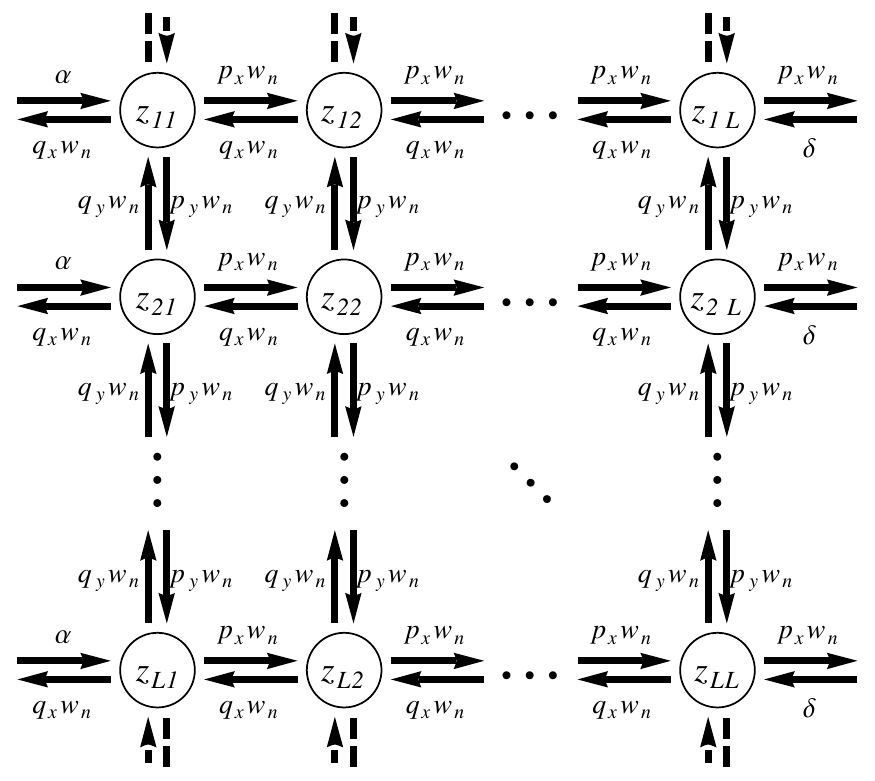}
	\end{center}
	\caption{2-$d$ ZRP with anisotropic hopping rates. The input rates $\alpha$ and $\delta$ determine the particle reservoir densities in the $x$-direction. Periodic boundary conditions in the $y$-direction are assumed.}
	\label{LxL}
\end{figure}

We now present a test of the AFR (\ref{FinalFR}) for an anisotropic zero-range process (ZRP) on an $L\times L$ square lattice. The ZRP is a paradigmatic non-equilibrium model where the on-site interaction of particles can be tuned to model different physical scenarios. Specifically, each site may be occupied by any number of particles, the top-most of which jumps randomly to a neighbouring site after an exponentially-distributed waiting time. Figure~\ref{LxL} shows the transition rates: they are determined by an interaction factor $w_n$, which depends only on the number $n$ of particles on the departure site, multiplied by hopping rates for the different jump directions and boundaries. Here we choose symmetric hopping rates, $p_x=q_x$ and $p_y=q_y$, which implies that the system scales as a diffusive process in the hydrodynamic limit ($L\rightarrow\infty$). We also take boundary rates in the $x$-direction corresponding to reservoir densities $\rho_L$ and $\rho_R$, with $\rho_L>\rho_R$, to induce a rightwards mean current. A feature of this model is that, depending on the choice of the term $w_n$, the system may show a condensation phase transition where particles accumulate on one or more sites~\cite{levine2005}. Indeed, even with a well-defined steady state, a bounded $w_n$ such that
\begin{equation}
\lim_{n \to \infty} w_n = a <\infty
\end{equation} 
results in instantaneous condensation in regimes of large current fluctuations~\cite{harris2005}.

\section{ZRP--Hydrodynamic limit}

The mobility and diffusion coefficients of the ZRP in each direction are $\sigma_k=p_k z(\rho)$ and $D_k=p_k z'(\rho)$~\cite{masi1984a,kipnis1999a,bertini2002}. Here $z(\rho)$ is a fugacity parameter connected to the density by
\begin{equation}
\rho=z\frac{\partial}{\partial z}\log\mathcal{Z},
\end{equation}
where $\mathcal{Z}$ plays the role of a grand canonical partition function. The form of $\mathcal{Z}$ depends on $w_n$; for example, choosing $w_n=w$, we have
\begin{equation}
\frac{\sigma_x}{p_x}=\frac{\sigma_y}{p_y}=w\frac{\rho}{\rho+1}
\end{equation}
and
\begin{equation}
\frac{D_x}{p_x}=\frac{D_y}{p_y}=\frac{w}{(\rho+1)^{2}}. 
\end{equation}
In this case, we can explicitly minimise Eq.~(\ref{SimplRF}) to find the optimal density profile as a solution of the associated Euler-Lagrange equation
\begin{equation}
\sum\limits_{k=1}^{2}\frac{2 D_k (\rho_{x_k}^{(1)})^2 \partial_\rho D_k+2 D^2_k \rho^{(2)}_{x_k}}{4 \sigma_k}-\frac{\left(D^2_k \rho^2_{x_k}-J^2_k\right) \partial_\rho \sigma_k}{4\sigma^2_k}=0, \label{SecondOrderPDE-Profile}
\end{equation}
where we have used the notation $\rho_{x_k}^{(n)}=\partial^n\rho/\partial x_k^n$.

\begin{figure}[t]
	\begin{center}
		\includegraphics[width=3.4in]{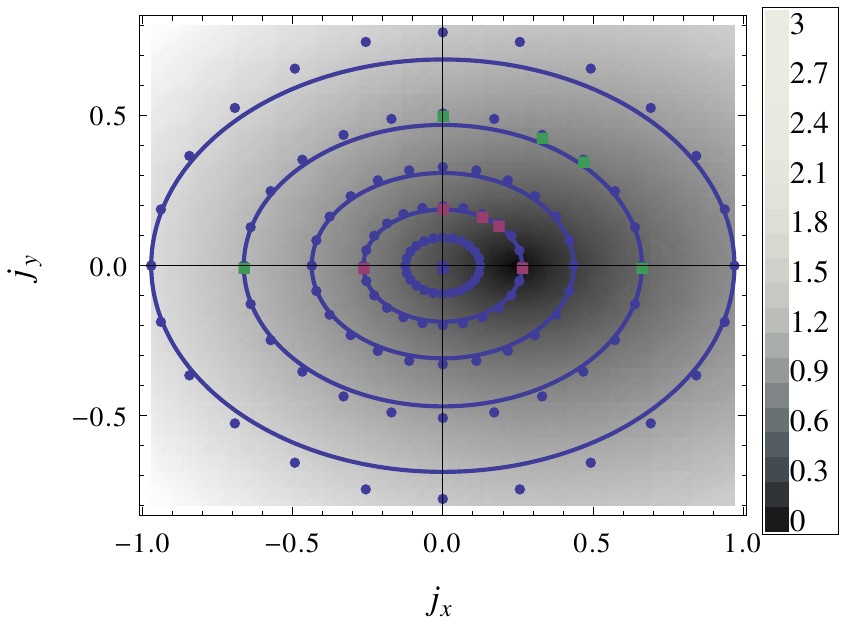}
	\end{center}
	\caption{Current rate function for the ZRP with interaction $w_n=1$ and boundary densities $\rho_L=1/2$ and $\rho_R=1/10$ (leading to $\mb{E}$ in the positive $x$-direction). Shaded background: Magnitude of $\hat e(\mb{J})$ calculated in the hydrodynamic approach (darkest at minimum). Solid lines: Currents satisfying the AFR as determined by the ellipse equation (\ref{eqell1}). Blue circles: Results from the microscopic solution of the model for $L=10^5$ with points of constant $e_L(\mb{\lambda})$ mapped by Legendre transform to the current space. Coloured squares: Points selected for Fig.~\ref{DensProfCurr}.}
	\label{DensPloRF}
\end{figure}

Following the established procedure in~\cite{perez-espigares2011a} we argue that the optimal density profile has no structure in the $y$-direction due to periodic boundary conditions. This allows us to equate the terms $2D_k^2 \rho^{(2)}_{x_k}$ and $D_k^2 \partial_\rho (\rho_{x_k}^{(1)})^2$. Integrating (\ref{SecondOrderPDE-Profile}), we then find that the optimal density profile is given by the first-order differential equation
\begin{equation}
	\left( D\nabla\rho \right)^{T}\sigmas\left( D\nabla\rho \right) = \mb{J}^{T}\sigmas\mb{J} + 4C, \label{MinDens}
\end{equation}
where $C$ is a constant of integration. Thus, we can explicitly find the optimal density profile and the current rate function. A similar calculation can be done for $w_n=n$, which corresponds to non-interacting particles. In both cases, the results confirm that the currents satisfying the AFR (\ref{FinalFR}) are located on ellipses verifying (\ref{eqell1}). This is shown for the interacting case $w_n=1$ in Fig.~\ref{DensPloRF} with hopping rates $p_x=1$ and $p_y=1/2$. Moreover, the optimal density profile associated with currents on each ellipse is invariant, a non-trivial result which follows again from~(\ref{MinDens1}).

\begin{figure}[t]
	\begin{center}
		\includegraphics[width=3.4in]{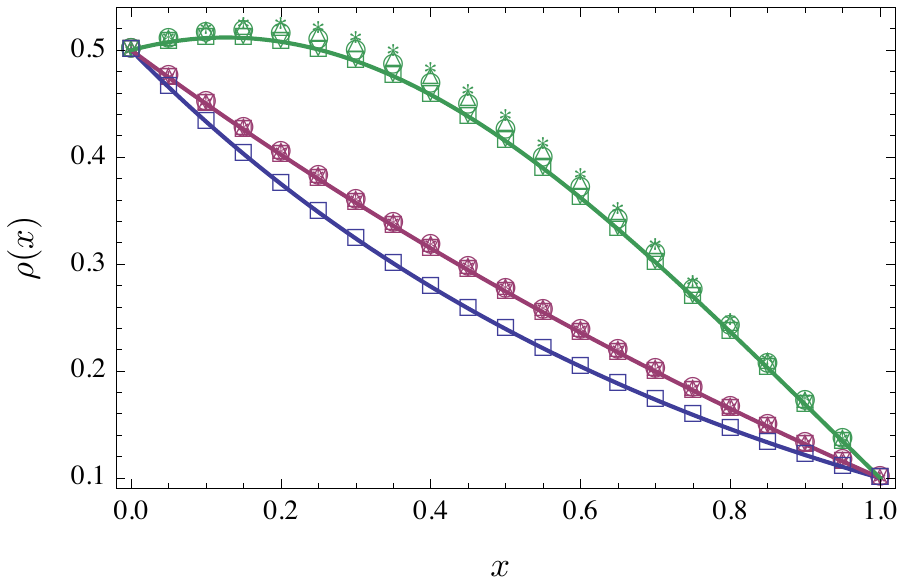}
	\end{center}
	\caption{Projection of the optimal density profiles in the $x$-direction for the currents on ellipses of Fig.~\ref{DensPloRF}. From bottom to top, data correspond to currents on ellipses with $x$-intercept $\left\{(0,0),(0.264,0),(0.663,0)\right\}$. Solid lines: Hydrodynamic theory. Symbols: Microscopic results for $L=10^5$ and the current values marked in Fig.~\ref{DensPloRF} with the same colours. The symbols for the angles with the $x$-axis are: 0 ($\square$), $\pi/4$ ($\ocircle$), $\pi/3$ ($\bigtriangleup$), $\pi/2$ ($*$), $\pi$ ($\bigtriangledown$).}
	\label{DensProfCurr}
\end{figure}

The specific shape of the optimal density profile depends on the current fluctuation considered, as shown in Fig.~\ref{DensProfCurr}, again for the case $w_n=1$. We observe that the non-linearity of Eq.~(\ref{MinDens}) results in two different kinds of density profile: for small fluctuations, the density is maximal at the left boundary and decreases monotonically, whereas for large fluctuations, the maximum density occurs at a point $x_{\text{max}}$ between the two boundaries. For $w_n=w$, the density at $x_{\text{max}}$ diverges at a critical current given by $\mb{J}_c^T\Lambda \mb{J}_c=K$ where $K$ can be explicitly calculated. This is the signature of the condensation mentioned above. The present analysis does not address the probability of current fluctuations outside this ellipse; indeed, as for the GCFR and IFR, we do not expect our AFR to hold in this regime, cf.~\cite{harris2006}.

\section{ZRP--microscopic approach}

Remarkably, it is possible to obtain exact results for the fluctuations of the current $\mb{J}$ in the ZRP for any lattice size and any interaction $w_n$, providing a precise test of the AFR and the validity of the assumptions behind it. These results follow by calculating the so-called \emph{scaled cumulant generating function} (SCGF),
\begin{equation}
	e_L(\mb{\lambda})=\lim\limits_{t\rightarrow\infty}-\frac{1}{tL^{d}}\log\langle \exp\left(-tL^d\mb{\lambda}\cdot\mb{J}\right)\rangle,
\label{SCGF}
\end{equation}
where $\langle\cdot\rangle$ denotes the expectation value. Note that the SCGF is here defined for systems of finite $L$; the corresponding hydrodynamic quantity with $L\to\infty$ is denoted by 
$e(\mb{\lambda})$.

The finite-$L$ SCGF can be explicitly calculated by writing the Master equation analogously to a quantum Schr\"odinger equation~\cite{schutz2001a} and extracting $e_L(\mb{\lambda})$ as the lowest eigenvalue of some modified Hamiltonian. The optimal density profile is then obtained from the corresponding eigenvector. As a higher dimensional generalization of the analysis in~\cite{harris2005}, we argue that for our system this eigenvector has a product form and, in practice, the calculation then involves solving an $L\times L$ system of linear equations for modified fugacities as a function of $\mb{\lambda}$. From $e_L(\mb{\lambda})$, we can verify the AFR either by obtaining the finite-$L$ rate function $\hat e_L(\mb{J})$ as the Legendre transform of $e_L(\mb{\lambda})$ or by noticing that Eq.~(\ref{FinalFR}) translates into the symmetry,
\begin{equation}
	e(\mb{\lambda}+\mb{E})=e(\mb{\lambda}'+\mb{E}),
\label{SCGF_FR}
\end{equation}
where the vectors $\mb{\lambda}$ and $\mb{\lambda}'$ satisfy
\begin{equation}
\left(\mb{\lambda}-\mb{E}\right) ^{T} \sigma \left(\mb{\lambda} - \mb{E}\right) =\left(\mb{\lambda}'-\mb{E}\right) ^{T} \sigma \left(\mb{\lambda}'-\mb{E}\right).
\label{eqell2}
\end{equation}
Geometrically, this means that $e(\mb{\lambda})$ is constant for vectors $\mb{\lambda}$ located on ellipses around the field $\mb{E}$. These ellipses are related by Legendre transform to those seen for the current in Fig.~\ref{DensPloRF}.

Note that the modified fugacities involved in the microscopic solution have no explicit dependence on $w_n$ facilitating the solution for any form of realistic interaction. However, $w_n$ does control the relation between fugacity and density and, importantly, determines the current regime in which the SCGF is given by the calculated lowest eigenvalue and the fluctuation relation is expected to hold. For example, in the case $w_n=1$ we can numerically calculate the maximum current before condensation, finding a bound consistent with the hydrodynamic approach.

The result of the Legendre transform of the ellipses of constant $e(\mb{\lambda})$ for the case $w_n=1$ are shown as data points in Fig.~\ref{DensPloRF} for the largest size studied, $L=10^5$. There we see a good agreement between the microscopic and hydrodynamic results, especially for small current magnitudes and for angles near the forward and backwards currents (i.e., approximately along and opposite to the direction of $\mb{E}$), confirming the AFR for these currents. For large fluctuations perpendicular to the field, there are discrepancies between the AFR prediction of the hydrodynamic approach and the microscopic solution of the model, which are discussed in more detail below.

Using the microscopic solution of the model, we can also examine the underlying structure of the optimal density profiles. In Fig.~\ref{DensProfCurr}, we compare the resulting optimal density profiles projected in the $x$-direction for increasing currents using microscopic and hydrodynamic theories. For $10^5 \times 10^5$ sites, the density profiles associated with forward and backward currents match exactly with the hydrodynamic result, whereas for large fluctuations and angles close to $\pi/2$, there are some deviations, which reflect at the level of density the differences seen in the rate function.

\begin{figure}[t]
	\begin{center}
		\includegraphics[width=3.4in]{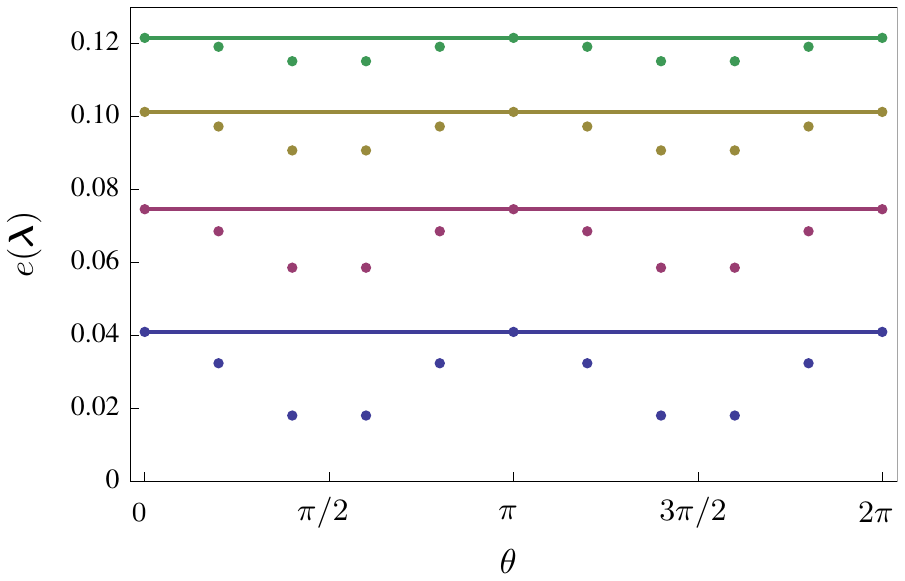}
	\end{center}
	\caption{SCGF $e(\mb{\lambda})$ of the ZRP ($w_n=n$, boundary conditions $\rho_L=1/2$, and $\rho_R=1/10$) as a function of the angle between ~$\mb{\lambda}-\mb{E}$ and $\mb{E}$. The different curves correspond to increasing ellipses around $\mb{E}$ (from top to bottom). Data points: Microscopic results obtained for the largest size studied, $L=10^5$.}
	\label{AFRmicromacroGF}
\end{figure}

To study these differences in more detail, we show in Fig.~\ref{AFRmicromacroGF} the SCGF $e(\mb{\lambda})$ for different values of $\mb{\lambda}$ on ellipses around the field as a function of the angle $\theta$ between $\mb{\lambda}-\mb{E}$ and the conjugate field $\mb{E}$ (i.e., the azimuthal angle measured from the ellipse principal axis in the field direction). For illustrative purposes we choose now the non-interacting case $w_n=n$ but we have checked that the behaviour is the same for $w_n=1$ in the absence of condensation. As for the rate function and density profiles, we see that the value of the SCGF, obtained from the microscopic solution of the ZRP, matches the constant hydrodynamic prediction of Eq.~(\ref{SCGF_FR}) for angles close to $0$ and $\pi$ and for small current values, which correspond to larger values of the SCGF. However, for large currents and angles close to $\pi/2$ the two approaches differ, which means that the AFR is not exactly satisfied in this regime.

\begin{figure}[t]
	\begin{center}
		\includegraphics[width=3.4in]{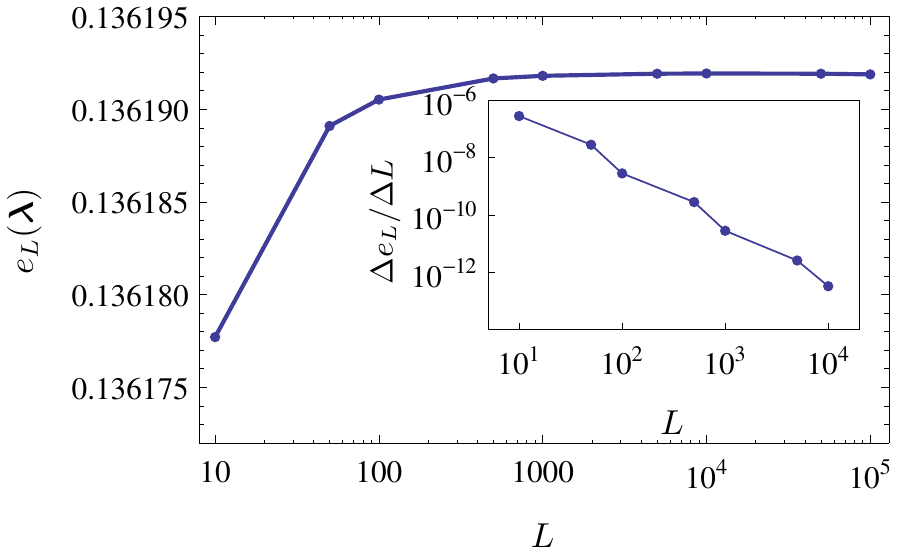}
	\end{center}
	\caption{Log-linear plot of the dominant eigenvalue $e_L$ of the modified Hamiltonian as a function of $L$. Results correspond to the ZRP with $w_n=n$ for the fluctuation $\mb{\lambda}=\left(0.534,-0.033\right)$ (i.e., current $\mb{J}=\left(0.249,0.015\right)$). Inset: Log-log plot of the discrete derivative $\Delta e_L/\Delta L$, showing that the slope of $e_L$ goes to zero for increasing $L$. Lines provided as a guide to the eye.}
	\label{loglogplot}
\end{figure}

For the original IFR, similar discrepancies at the level of the SCGF were seen in simulations of the KMP-process for system size $L=32$~\cite{hurtado2011} and were interpreted as a finite-size effect. For the ZRP, this cannot be the case: as shown in Fig.~\ref{loglogplot}, the dominant eigenvalue $e_L(\mb{\lambda})$ of the $L\times L$ linear system that we solve converges quickly in the limit $L\ra\infty$. In fact, for $L=100$, the dominant eigenvalue is already very close to its converging value. This is shown in Fig.~\ref{loglogplot} for one particular value of $\mb{\lambda}$; however, we have checked that all values reported in Fig.~\ref{AFRmicromacroGF} converge in the same way, which means that the data reported in that figure for $L=10^5$ are a clear indication that the AFR holds, as stated before, in the regime of small currents, as well as along and opposite the field $\mb{E}$, but only approximately for large currents perpendicular to $\mb{E}$.

At this point, we do not have a complete explanation of the difference between the hydrodynamically-predicted AFR and the microscopic solution of the ZRP.  However, we believe the reason is that although all three assumptions (\emph{ii})--(\emph{iv}) are necessary for the hydrodynamic derivation, assumption (\emph{iv}) is not exactly satisfied by the ZRP.   Significantly, this last assumption of spatially homogeneous optimal current profiles is not made in the microscopic analysis; we implicitly use only assumptions (\emph{ii}) and (\emph{iii}) which together imply the weaker condition that the optimal current is divergence free and has no structure perpendicular to the field, i.e., in the $y$-direction here. This leaves open the possibility that the $y$-component of the current has an $x$-dependence in violation of assumption (\emph{iv}) -- indeed for models (such as the ZRP) with a density-dependent mobility this might be expected whenever the optimal density profile is $x$-dependent. This argument is supported by the observation that, when the optimal density profile is approximately constant (e.g., close to equilibrium $\rho_L=\rho_R$) so that the mobility is approximately homogeneous throughout the lattice, the microscopic results do seem to converge towards the hydrodynamically-predicted AFR. A more detailed analysis of this property will be presented in a future paper \cite{villavicencio2014}.

\section{Conclusion}

We have presented in this paper an extension of the recently-introduced IFR to diffusive systems having anisotropic driving rates. This AFR shows very good agreement with exact microscopic calculations for small current fluctuations, as well as for large currents close to the driving field. This is particularly relevant, since it allows the symmetry to be efficiently probed experimentally (say, for manipulated Brownian particles with anisotropy) without the need to measure rare backward fluctuations. Moreover, from a theoretical point of view, the observation that microscopic results for perpendicular current fluctuations in a specific model may not converge to the AFR sheds light on the underlying physical assumptions (also required for the original IFR). The exactly solvable ZRP provides an ideal playground for future work focusing on the precise role of these assumptions.
 
In the wider framework of non-equilibrium statistical mechanics, we note that the IFR itself has important consequences for current cumulants and non-linear response coefficients, leading indeed to a hierarchy of relations beyond Onsager and Green-Kubo results~\cite{hurtado2011}. We anticipate a similarly rich vein of possible implications from the AFR.  Furthermore, although our AFR was derived within the context of diffusive systems, we believe that an analogous argument holds more generally for a broader class of models with local conservation laws (e.g., systems of Ginzburg-Landau type) meaning that the results presented here have potential relevance for many anisotropic processes of interest.

\begin{acknowledgements}
RVS is supported by the Mexican National Council for Science and Technology (CONACyT) Scholarship scheme. RJH and HT are grateful to the Kavli Institute for Theoretical Physics China, for support and hospitality while this work was being finalized. 
\end{acknowledgements}

\bibliographystyle{eplbib}  
\bibliography{masterbibmin}
\end{document}